\documentstyle[aps,preprint,tighten]{revtex}

\begin{document}
\preprint{KSUCNR-009-94}
\draft
\title{Heavy resonance production in high energy nuclear collisions}
\author{David Seibert\thanks{Electronic mail (internet):
seibert@scorpio.kent.edu.} and George Fai\thanks{Electronic mail (internet):
fai@ksuvxd.kent.edu.}}
\address{Center for Nuclear Research\\ Department of Physics, Kent State
University, Kent, OH 44242}
\date{July 7, 1994}
\maketitle
\begin{abstract}
We estimate freezeout conditions for $s$, $c$, and $b$ quarks in
high energy nuclear collisions.  Freezeout is due either to loss of thermal
contact, or to particles ``wandering'' out of the region of hot matter.
We then develop a thermal recombination model in which both single-particle
(quark and antiquark) and two-particle (quark-antiquark) densities are
conserved.  Conservation of two-particle densities is necessary because
quarks and antiquarks are always produced in coincidence, so that the local
two-particle density can be much larger than the product of the
single-particle densities.  We use the freezeout conditions and
recombination model to discuss heavy resonance production at zero baryon
density in high energy nuclear collisions.
\end{abstract}

\section{Introduction}

One of the striking features of high energy proton-proton, proton-nucleus
and nucleus-nucleus collisions is the production of heavy resonances.
Since these resonances appear as a consequence of raising the collision
energy, it is natural to ask, ``Can heavy resonances be used experimentally
to signal a deconfined phase of matter?'' Enhanced strangeness production
was in fact proposed as one of the first potential signals of the Quark
Gluon Plasma (QGP).\cite{rstrange}  J/$\psi$  production \cite{rJ/psi}
can be looked upon as changing the focus to the heavier charm ($c$) quark
from the much lighter strange ($s$) quark.  Here we aim at a unified
treatment of heavy quark composites, extending the scope to include also
the bottom ($b$) quark.

In view of the very different masses of the $s$, $c$, and $b$ quarks, and
of the mesons that contain these quarks, it is reasonable to
assume that the various flavors will stop interacting with the
expanding collision system at different times and thus carry ``frozen out''
information about different stages of the time evolution, reflecting
their last interactions. This calls for a detailed examination of the
freezeout conditions for each flavor. The concept of different freezeout
times for different species is well known in studies of the Early Universe,
and has also been discussed in the context of nuclear collisions at lower
energies.\cite{rn}

Our physical picture assumes thermal equilibrium, and chemical equilibrium
between resonances with the same heavy quark content (but not between heavy
quarks and antiquarks) at high temperature.  We allow resonances to
depart from chemical equilibrium after freezeout.  In determining the
freezeout chemical potentials we pay particular attention to the {\it local}
quark-antiquark ($Q\overline{Q}$) pair density.  Since quarks and antiquarks
are always created in pairs, the local pair density may be significantly
higher than the product of the single-particle densities.  We take this
important effect into account in terms of a pair chemical potential (see
Section~\ref{trm}).  This pair chemical potential is the main difference
between our recombination model and others that have been used recently to
study strange \cite{rrcss,rlthsr} and heavier exotic \cite{rzl}
resonances.

In Section~\ref{fc} we discuss two conditions which may signal that
heavy resonances cease to interact with the hot matter, and estimate
freezeout temperatures and proper times for $s$, $c$, and $b$ quark
resonances in selected $AA$ collisions at CERN's Super Proton Synchrotron
(SPS, for which $\sqrt{s}=20$ GeV/nucleon) and future Large Hadron Collider
(LHC, $\sqrt{s}=7$ TeV/nucleon), and at Brookhaven's future Relativistic
Heavy Ion Collider (RHIC, $\sqrt{s}=200$ GeV/nucleon).  We obtain a rich
variety of freezeout scenarios, depending on the properties of the
transition from the hadronic phase to the high temperature (deconfined)
phase. We also calculate the mean distance a heavy quark moves before
freezeout and the mean number of interactions it has; the mean number
of interactions is high enough that thermal recombination models can
be expected to work for heavy ion collisions at or above SPS energy,
but not necessarily for collisions of lighter ions.  In Section~\ref{trm}
we describe our thermal recombination model. We then give an illustrative
application to J/$\psi$ suppression, and calculate freezeout densities
and the fractions of $c$ and $b$ quarks that end up in $c\overline{c}$
and $b\overline{b}$ resonances.  Finally, in Section~\ref{sc} we
summarize our work.

\section{Freezeout conditions} \label{fc}

In this section, we estimate the freezeout conditions for the different
heavy quark flavors.  We consider two mechanisms for freezeout.  In
subsection~\ref{sscf} we discuss freezeout due to loss of thermal
contact, which we refer to as ``contact'' freezeout.  In
subsection~\ref{sswf} we discuss freezeout by emission, which we estimate
by treating the quarks as random walkers, and hence refer to as
``wandering'' freezeout.  We then summarize the results of the two
freezeout scenarios in subsection~\ref{ssfs}.

The $s$ quark mass, $m_s$, is approximately equal to the quantum
chromodynamics (QCD) transition temperature, $T_c \approx 150-200$ MeV,
while the $c$ and $b$ quark masses, respectively $m_c$ and $m_b$, are much
greater.  [We use $m_s=0.2$ GeV, $m_c=1.5$ GeV, and $m_b=5$ GeV.]  Thus,
freezeout conditions for $c$ and $b$ quarks are slightly different than
those for $s$ quarks.  Because of this, we estimate freezeout conditions
under each scenario for $c$ and $b$ quarks together, and then for $s$
quarks.

\subsection{Loss of thermal contact} \label{sscf}

One possibility is that freezeout occurs when the thermal contact between
the heavy quark and the hot matter ends.  In this case, the temperature at
freezeout, $T_f$, is determined by setting the mean free path equal to
the mean distance to leave the hot matter.  If the hot matter forms an
infinite cylinder of radius $R_f$ at freezeout, the initial location is
equally likely to be anywhere within the cylinder, and any initial
direction of motion is equally likely, the mean distance to leave the hot
matter is $4R_f/3$.  The thermal heavy quark mean free path is
\begin{equation}
\lambda_Q(T) = \frac {v_Q(T)} {\Gamma_Q(T)},
\end{equation}
where $v_Q$ is the velocity and $\Gamma_Q$ is the collision rate.
For heavy quarks in deconfined matter, we take
\begin{eqnarray}
v_Q(T) &=& \left( \frac {3T} {m_Q} \right)^{1/2}, \label{evQ} \\[12pt]
\Gamma^{(d)}_Q(T) &=& \frac {4} {3\pi} \, T, \label{eGQd} \\[12pt]
\lambda^{(d)}_Q(T) &=& \frac {3\pi} {4} \sqrt{3 / m_Q T}, \label{elQd}
\end{eqnarray}
where we use $\Gamma_Q$ from Ref.~\cite{rPis} with $N=3$ colors,
Casimir factor $C_f=4/3$, $N_f=3$ light flavors ($u$, $d$, $s$) and
strong coupling constant $g=2$.  We have dropped terms of order
$v_Q \ln (1/v_Q)$ in these equations, as these are small and do not
affect our results much.

In the coexistence region, the mean free path is
\begin{equation}
\lambda^{(c)}_Q = \frac {\lambda^{(d)}_Q(T_c) \, \lambda^{(h)}_Q(T_c)}
{(1-x) \lambda^{(d)}_Q(T_c) \, + \, x \lambda^{(h)}_Q(T_c)},
\end{equation}
where $x$ is the fraction of matter in the deconfined phase.  We
estimate mean free paths in hadronic matter at $T_c$ by assuming that
the density of scatterers is proportional to the entropy density in the
two phases, and that cross sections remain constant:
\begin{equation}
\lambda^{(h)}(T_c) = \frac {\nu} {2} \lambda^{(d)}(T_c),
\end{equation}
where $\nu = s_d(T_c)/s_h(T_c)$ is the ratio of entropy densities in the
two phases at $T_c$.  This is equivalent to assuming that the number of
scatterers is proportional to the entropy, and that the phase transition
dilutes the scatterers without affecting their cross sections; the extra
factor of $1/2$ occurs because quarks must travel in mesons (which we
treat as two quarks) in the hadronic phase.  We thus obtain
\begin{equation}
\lambda^{(c)}_Q = \frac {\nu}
{2 \, + \, (\nu-2) x} \, \lambda^{(d)}_Q(T_c). \label{elc}
\end{equation}

For hadronic gas, we expect the mean free path to be
\begin{equation}
\lambda^{(h)}_Q(T) = \frac {v_Q(T)} {\rho \sigma_Q},
\end{equation}
where $\rho$ is the density of light particles (with $m \ll T$) and
$\sigma_Q$ is the mean cross section.  The thermal velocity is approximately
the same for heavy quarks and mesons containing heavy quarks,
so at $T_c$ we find
\begin{equation}
\sigma_Q = \frac {72}{\pi^3} \, \nu^{-1} \, T_c^{-2}
 \simeq \frac {2.3} {\nu} \, T_c^{-2},
\end{equation}
assuming continuity of $\lambda_Q$, and that $\rho$ is given by the density
of a massless pion gas,
\begin{equation}
\rho = \frac {\pi^2} {27} T_c^3.
\end{equation}
For $T_c = 150-200$ MeV and $\nu = 5-10$, $\sigma_Q = 2-8$ mb, in
reasonable agreement with currently used values for heavy meson
cross-sections.  However, different heavy quark mesons may have very
different mean free paths, so it is not clear that all meson species
containing any given heavy quark flavor will freeze out at the same time.
This issue is beyond the scope of this paper, so we leave it for future
work.

We extrapolate to lower temperatures by assuming that
$\sigma_Q$ is constant and $\rho \propto T^3$, so
\begin{equation}
\lambda_Q^{(h)}(T) = \frac {\nu} {2} \lambda_Q^{(d)}(T_c)
\left( \frac {T_c} {T} \right)^{5/2}. \label{elQh}
\end{equation}
This estimate of $\lambda_Q^{(h)}$ is reasonably accurate near $T_c$.
However, for $T \ll m_{\pi}$ the mean free path will grow exponentially
with $T^{-1}$ as the density drops, so freezeout will occur before $T$
drops much below $m_{\pi}$.

For freezeout occurring from deconfined matter, we obtain
\begin{equation}
\frac 4 3 R_f = \frac {3\pi} {4} \sqrt{3 / m_Q T_f} \, \rightarrow \,
T_f = \frac {243 \pi^2} {256} \, m_Q^{-1} \, R_f^{-2}
\simeq 9 \, m_Q^{-1} \, R_f^{-2}. \label{eTfd1}
\end{equation}
Freezeout occurs for $T \geq T_c$ if
\begin{equation}
R_f ~\leq~ \frac {9 \pi \sqrt{3}} {16}  \,
\left( m_Q T_c \right)^{-1/2} \simeq 3 \left( m_Q T_c \right)^{-1/2}.
\end{equation}
``Contact'' freezeout occurs from the coexistence region as long as
\begin{equation}
R_f \leq \frac {9 \pi \sqrt{3}} {32}  \, \nu \,
\left( m_Q T_c \right)^{-1/2} \simeq 1.5 \, \nu \,
\left( m_Q T_c \right)^{-1/2},
\end{equation}
and occurs from normal hadronic matter for larger values of $R_f$.
Note that in all equations above $R_f$ can depend on $\tau_f$ or $T_f$,
so that the equations are valid even with transverse expansion of the hot
matter as long as $R_f$ is treated properly.

The $s$ quark mass is small, so its velocity can be taken to be unity, and the
mean free path is
\begin{equation}
\lambda_s^{(d)}(T) = \frac {1} {\Gamma_s^{(d)}(T)} = 2 \, T^{-1},
\label{elsd} \end{equation}
where we obtain $\Gamma_s^{(d)}$ for light quarks from Ref.~\cite{rPis}.
We then find that ``contact'' freezeout of $s$ quarks from deconfined
matter occurs when
\begin{eqnarray}
T_f = \frac 3 2 R_f^{-1}. \label{eTsd1}
\end{eqnarray}
Freezeout thus occurs at or above $T_c$ if
\begin{equation}
R_f \leq \frac 3 2 T_c^{-1}.
\end{equation}

We estimate the mean free path for an $s$ quark in the coexistence
region just as we did for the $c$ and $b$ quarks.  The thermal velocity
in hadronic matter is still approximately unity near $T_c$, as the mass
of the lightest meson containing an $s$ quark is not too much greater
than the transition temperature.  If we again assume that the number of
scatterers is proportional to the entropy, and cross sections are
constant, we obtain Eq.~(\ref{elc}) as before.  This is equivalent to
assuming a cross section
\begin{equation}
\sigma_s = \frac {27/\pi^2} {\nu} \, T_c^{-2} \simeq 3-10 \, \mbox{mb},
\end{equation}
which again is not an unreasonable value.  We find that ``contact''
freezeout of $s$ quarks in the coexistence region occurs if
\begin{equation}
R_f \leq \frac 3 4 \, \nu \, T_c^{-1}.
\end{equation}
For larger values of $R_f$, the $s$ quarks will freeze out from normal
hadronic matter.

\subsection{Emission (``wandering'')} \label{sswf}

A second possibility is that freezeout occurs when the quarks leave the hot
matter.  This happens when the path length becomes equal to the mean distance
to leave the cylinder.  We calculate the path length by treating the quarks
as random walkers, with thermal velocities and mean free paths.  The rms
distance moved by a quark (or antiquark) of flavor $Q$, $r_Q$, from creation
at proper time $\tau_0 \simeq 0$ to freezeout at proper time $\tau_f$ is
given by
\begin{equation}
r_Q^2 = \int_0^{\tau_f} d\tau \, v_Q[T(\tau)] \, \lambda_Q[T(\tau)],
\label{er2Q} \end{equation}
where $v_Q$ and $\lambda_Q$ are respectively the thermal velocity and mean
free path for flavor $Q$, evaluated at $T(\tau)$.  We relate $T$ and $\tau$
by treating the deconfined matter as a non-interacting gas with 16 boson and
24 fermion degrees of freedom, assuming that entropy is
conserved,\cite{rBj} and that the final state can be treated as a
massless Bose gas, so that
\begin{equation}
\pi R_f^2 \, \tau \times \, \frac {74 \pi^2} {45} \, T^3 ~=~
3.6 \, dN/dy, \label{eTtau}
\end{equation}
where $dN/dy$ is the total rapidity density (charged plus neutral).  As we
neglect transverse expansion, we set $R_f$ equal to the nuclear radius,
\begin{equation}
R_f = 1.2 \, A^{1/3}~\mbox{fm}.
\end{equation}
The assumption that $\tau_0=0$ is good for $c$ and $b$ quarks, which are
only created early in the collision, but not necessarily for $s$ quarks
which are created as long as there is any deconfined matter.\cite{ras}
However, we make this approximation for simplicity.

``Wandering'' freezeout of heavy quarks occurs in deconfined matter when
\begin{equation}
\left( \frac 4 3 R_f \right)^2 \leq \frac {9\pi} {4 m_Q} \, \tau_c \,
\rightarrow \, dN/dy \geq \frac {37 \times 64 \pi^2} {3^8} \,
m_Q \, T_c^3 \, R_f^4 \simeq 3.6 \, m_Q \, T_c^3 \, R_f^4,
\label{edNdyd2} \end{equation}
where $\tau_c$ is the proper time at which the deconfined matter reaches
$T_c$.  Including transverse expansion increases $R_f$ and decreases $T$
for fixed $\tau$, making it less likely that $c$ or $b$ quarks freeze out
before the hot matter reaches $T_c$.  The condition for ``wandering''
freezeout in the coexistence region is
\begin{equation}
dN/dy \geq \frac {2368 \pi^2} {6561
\left\{ 1 + (\nu-1) \left[ (\nu-1)-(\nu-2)\ln 2 \right] \right\}} \,
m_Q \, T_c^3 \, R_f^4,
\end{equation}
if there is no transverse expansion.  For smaller values of $dN/dy$,
``wandering'' freezeout occurs during the hadronic phase.

For ``wandering'' of $s$ quarks, we again take Eq.~(\ref{er2Q}), but with
$v_Q=1$.  Using $\lambda_s(T)$ from Eq.~(\ref{elsd}), we find that $s$
quarks freeze out from deconfined matter if
\begin{equation}
dN/dy \geq \frac {1184 \pi^3} {2187} \,
T_c^4 \, R_f^4 \simeq 16 \, T_c^4 \, R_f^4.
\end{equation}
``Wandering'' freezeout from the coexistence region will occur if
\begin{equation}
dN/dy \geq \frac {1184 \pi^3} {729
\left\{ 3 + 4 (\nu-1) \left[ (\nu-1)-(\nu-2)\ln 2 \right] \right\} } \,
T_c^4 \, R_f^4.
\end{equation}
For smaller values of $dN/dy$, ``wandering'' freezeout takes place from
hadronic matter. Again, including transverse expansion would shorten the
duration of the hot matter at any entropy density, so the above inequalities
give solid lower limits on the values of $dN/dy$ needed for ``wandering''
freezeout.

\subsection{Freezeout summary} \label{ssfs}

We have evaluated freezeout conditions for $s$, $c$, and $b$ quarks
in proton collisions at SPS, RHIC, and LHC energies.  We find that the mean
number of collisions before freezeout,
\begin{equation}
n_f = \int_0^{\tau_f} d\tau \, \Gamma \left[ T(\tau) \right],
\end{equation}
is much less than unity for proton collisions at any feasible energy.
Thus, the behavior of heavy quarks in proton collisions probably cannot
be described by the quasi-equilibrium models used here, and so
predictions for proton collisions are beyond the scope of this paper.

In Table~I, we show freezeout conditions for the various heavy quark
flavors in selected $AA$ collisions at SPS, RHIC, and LHC.
We use four scenarios, with $\nu=5$ and 10, and $T_c=150$ and 200 MeV.
This gives a reasonable range of values for $T_c$, while $\nu=10$ is near
the upper limit for a strong first-order QCD phase transition, and
$\nu=5$ gives a moderately strong phase transition.  These values are
compatible with lattice gauge theory results.

The number of collisions per quark is only weakly dependent on $\nu$,
$T_c$, and flavor.  For S+S collisions at SPS, $n_f \simeq 1$, so
quasi-equilibrium models are not ridiculous but are also not very well
justified.  For Au+Au collisions, $n_f$ is significantly larger than unity,
and thus quasi-equilibrium models are at least self-consistent even at SPS.
This is in general agreement with Parton Cascade Model (PCM)
results for Au+Au collisions at RHIC and LHC,\cite{rgkce} which show that
all quarks follow approximately hydrodynamic behavior, with mean momenta
following the temperature of the surrounding matter.

None of the quark flavors freeze out above $T_c$.  The $s$ quarks freeze out
before the end of coexistence in any collision if $\nu=10$.  For $\nu=5$,
they may freeze out in the hadronic phase, especially if $T_c$ is relatively
high.  In general, freezeout of $s$ quarks typically occurs at $T_f \simeq
T_c$.

The $c$ quarks also freeze out in the coexistence region if $\nu=10$, except
in heavy ion collisions at SPS and RHIC if $T_c$ is large.  If
$\nu=5$, they freeze out below $T_c$.  Since these are upper limits, it is
likely that $c$ quarks also freeze out in the coexistence region if there is
a strong QCD phase transition, but they freeze out from hadronic matter
if the transition is weak.

The $b$ quarks may freeze out in the coexistence region in Au+Au collisions
at LHC, but only if the transition is strong and $T_c$ is relatively
low.  In all other cases considered, they freeze out far below $T_c$.  Thus,
accurate predictions of $b$ quark physics in Au+Au collisions below LHC
energy (and possibly at LHC, if $T_c \simeq 200$ MeV) probably
requires accurate cross sections for the various $b$ mesons in hadronic
matter, and possibly the use of a microscopic hadronic cascade code.

The energy required for freezeout in the coexistence region will be lower in
p$A$ and $AB$ collisions than in $AA$ collisions.  However, including
transverse expansion will raise the energy thresholds, as might more
realistic estimates of heavy resonance cross sections in hadronic matter, so
the values given here are probably lower bounds for freezeout before the end
of coexistence.

\section{The thermal recombination model} \label{trm}

We use a thermal recombination model that is similar to the models of
Refs.~\cite{rrcss,rlthsr,rzl}.  The major difference is that we introduce a
pair chemical potential to account for the fact that the quarks and antiquarks
are always produced in pairs, so the local density of pairs may be much
larger than the product of the quark and antiquark densities:
\begin{equation}
\rho^{(2)}_{Q\overline{Q}}(x,x) ~\gg~ \rho_Q(x) \, \rho_{\overline{Q}}(x).
\end{equation}
We assume that the heavy quark flavors ($s$, $c$, and $b$) are far from
chemical equilibrium before they stop interacting thermally,\cite{ras} so
that the densities of heavy quarks and antiquarks at thermal freezeout are
separately conserved.  The chemical potential for hadron species $n$ is
determined by the quark, antiquark, and pair chemical potentials,
respectively $\mu_Q$, $\mu_{\overline{Q}}$, and $\mu_{Q\overline{Q}}$:
\begin{equation}
\mu_n = \sum_Q \left( k^{(Q)}_n \mu_Q \,
+ \, k^{(\overline{Q})}_n \mu_{\overline{Q}}
+ \, k^{(Q\overline{Q})}_n \mu_{Q\overline{Q}}
\right). \label{emu}
\end{equation}
Here $k^{(Q)}_n$ ($k^{(\overline{Q})}_n$) is the number of quarks
(antiquarks) of flavor $Q$, and $k^{(Q\overline{Q})}_n$ is the number
of pairs (the smaller of $k^{(Q)}_n$ and $k^{(\overline{Q})}_n$).  We assume
that the number of hadrons of species $n$ produced is proportional to its
density at temperature $T_f$ and chemical potential $\mu_n$.

The chemical potentials are determined by conservation of quark and
antiquark densities, respectively $\rho_Q$ and $\rho_{\overline{Q}}$,
and the local two-particle quark-antiquark density,
$\rho^{(2)}_{Q\overline{Q}}$.  [We use the more compact notation
$\rho^{(2)}_{Q\overline{Q}} = \rho^{(2)}_{Q\overline{Q}}(x,x)$ for the
remainder of this paper.]  Since there is no chemical equilibrium between
quarks and antiquarks, the constraint $\mu_{\overline{Q}}=-\mu_Q$ does not
apply.  The equations are
\begin{eqnarray}
\rho_Q &=& \sum_{n_b} \frac {g_{n_b} k_{n_b}^{(Q)}} {2\pi^2}
\int_0^{\infty} \!\!\! \frac {p^2 \, dp} {e^{[E_{n_b}(p)-\mu_{n_b}]/T_f}-1}
+ \sum_{n_f} \frac {g_{n_f} k_{n_f}^{(Q)}} {2\pi^2} \int_0^{\infty}
\!\!\! \frac {p^2 \, dp} {e^{[E_{n_f}(p)-\mu_{n_f}]/T_f}+1},
\label{erhoQ} \\[12pt]
\rho_{\overline{Q}} &=& \sum_{n_b} \frac {g_{n_b}
k_{n_b}^{(\overline{Q})}} {2\pi^2} \int_0^{\infty} \!\!\!
\frac {p^2 \, dp} {e^{[E_{n_b}(p)-\mu_{n_b}]/T_f}-1} + \sum_{n_f}
\frac {g_{n_f} k_{n_f}^{(\overline{Q})}} {2\pi^2} \int_0^{\infty} \!\!\!
\frac {p^2 \, dp} {e^{[E_{n_f}(p)-\mu_{n_f}]/T_f}+1},
\label{erhoQbar} \\[12pt]
\frac {\rho^{(2)}_{Q\overline{Q}}}{\rho_Q} &=& \sum_{n_b} \frac {g_{n_b}
k_{n_b}^{(\overline{Q})}} {2\pi^2} \int_0^{\infty} \!\!\!
\frac {p^2 \, dp} {e^{[E_{n_b}(p)-\mu_{n_b}']/T_f}-1}
+ \sum_{n_f} \frac {g_{n_f} k_{n_f}^{(\overline{Q})}} {2\pi^2}
\int_0^{\infty} \!\!\! \frac {p^2 \, dp}
{e^{[E_{n_f}(p)-\mu_{n_f}']/T_f}+1};
\label{erhoQQbar} \\[12pt]
\mu_n' &=& \mu_n +\mu_{Q\overline{Q}}, \\[12pt]
E_n(p) &=& \left( p^2+m_n^2 \right)^{1/2}.
\end{eqnarray}
Here the sums are over bosons ($n_b$) and fermions ($n_f$), while $m_n$ and
$g_n$ are respectively the mass and degeneracy for hadron $n$.

We estimate $\rho^{(2)}_{Q\overline{Q}}$ by again treating the quarks as
random walkers.  We assume that the $Q$ and $\overline{Q}$ are distributed
randomly over a sphere centered on the point at which the pair was formed,
such that the rms distances from the center are $r_Q$.  We thus obtain
\begin{equation}
\rho^{(2)}_{Q\overline{Q}} \approx \rho_Q \, \left[ \rho_{\overline{Q}} \,
+ \, \frac {3} {4\pi} \left( \frac {3} {5 \, r_Q^2} \right)^{3/2} \right],
\label{erhoQQ} \end{equation}
where $\rho_Q$, $\rho_{\overline{Q}}$ and $r_Q$ are determined at
freezeout.  We obtain the rhs of Eq.~(\ref{erhoQQbar}) by averaging the
$Q\overline{Q}$ pairs in hadrons over the same sphere to get the mean value
of $\rho^{(2)}_{Q\overline{Q}}$ for the hadrons.   For copiously produced
quarks, $\rho^{(2)}_{Q\overline{Q}} \approx \rho_Q \rho_{\overline{Q}}$, so
the third constraint equation (\ref{erhoQQbar}) just gives
$\mu_{Q\overline{Q}} \approx 0$.

\subsection{J/$\psi$ suppression} \label{sJ/psi}

One possible application of this model is to understand the phenomenon
of J/$\psi$ suppression.  The prediction \cite{rJ/psi} is that
$c\overline{c}$ pairs will tend to form J/$\psi$ mesons less often in
collisions in which QGP is formed, as a result of
interactions of the pair with the deconfined matter.  Thus, if QGP is
formed in nuclear collisions, production of J/$\psi$ mesons will
be suppressed relative to production of $c\overline{c}$ pairs.

In this section, we show how suppression of heavy $Q\overline{Q}$ mesons
arises in our model.  For illustrative purposes, we consider a case for
which there are only three hadron species.  Species 1 hadrons contain a
single $Q$ and have degeneracy $g_1$ and mass $m_1$.  Species 2 hadrons are
the antiparticles of species 1, so they contain a single $\overline{Q}$ and
have degeneracy $g_1$ and mass $m_1$.  Species 3 hadrons contain a
$Q\overline{Q}$ pair, and have degeneracy $g_2$ and mass $m_2 \simeq 2m_1$.
This is similar to the situation for $c$ quarks, where the $D$ and
$\overline{D}$ could be regarded as species 1 and 2 and the J/$\psi$ as
species 3.

Because heavy quarks freeze out at $T_f \ll m_Q$, we use Maxwell-Boltzmann
statistics for the heavy hadrons.  We take $\rho_Q=\rho_{\overline{Q}}$,
because heavy quarks are always produced in pairs so the net density is
always zero.  We then solve the constraint equations, obtaining
\begin{equation}
\frac {\rho_3} {\rho_1} = \frac 1 2 \left\{ -1 \, + \,
\left[ 1 \, + \, \frac {4 g_2} {g_1^2} \left( \frac {2\pi m_2} {m_1^2 T_f}
\right)^{3/2} \, e^{(2m_1-m_2)/T_f} \, \frac {\rho^{(2)}_{Q\overline{Q}}}
{\rho_Q} \right]^{1/2} \right\}. \label{eQQex}
\end{equation}

It is clear from eq.~(\ref{eQQex}) that the relative densities of the
resonances depends only on $T_f$ and the local density,
$\rho^{(2)}_{Q\overline{Q}}/\rho_Q$.  This remains true for more complicated
resonance spectra as long as all resonances have at most one quark and one
antiquark, if the flavor is heavy enough to justify the use of
Maxwell-Boltzmann statistics.  These conditions are met by the flavors $c$
and $b$, so studies of the relative production of $c$ and $b$ resonances
(except for exotic multi-heavy hadrons \cite{rzl}) will be insensitive to
$\rho_c$ and $\rho_b$.  Thus, the only way to determine $\rho_c$ and
$\rho_b$ (in the absence of exotic multi-heavy hadrons) is through
comparisons of heavy and light resonance production.  Since there are many
multi-strange baryons, $\rho_s$ can be determined solely by studying strange
resonances, as long as these multi-strange baryons are detected.

\subsection{Freezeout density estimates}

We estimate freezeout densities from rapidity densities, staying as close
as possible to predictions supported by data.  In the absence of transverse
expansion,
\begin{equation}
\rho_Q = \rho_{\overline{Q}} = \frac {dN_Q/dy} {\pi \, R_f^2 \, \tau_f},
\end{equation}
and we obtain $\rho^{(2)}_{Q\overline{Q}}$ from eq.~(\ref{erhoQQ}).
The predicted freezeout densities are shown in Table~II; for the interested
reader, we explain our estimates of $dN_Q/dy$ below.  The main point of
interest is that $\rho^{(2)}_{Q\overline{Q}} \gg \rho_Q \rho_{\overline{Q}}$
for $b$ quarks in all collisions considered, and for $c$ quarks except in
Au+Au collisions at LHC.  For $s$ quarks, the difference is less than 10\%
in most collisions, which is probably within experimental error and thus not
of great interest.

The PCM reproduces strange particle production in p\=p collisions at
$\sqrt{s}=546$ GeV fairly well,\cite{rpcms} so we assume that it gives
reasonable predictions for $dN_s/dy$ in $AA$ collisions at RHIC and
LHC.\cite{rgscb}  The factor of 5 increase in $s$ production from RHIC to
LHC is roughly consistent with most $s$ quarks being produced thermally,
which gives $dN_s/dy \propto (dN/dy)^2/A^{2/3}$ for $AA$ collisions with
the same initial temperature.\cite{ras}  For S+S collisions at SPS, we
use data \cite{rSPS} to estimate $dN_s/dy$.  For Au+Au collisions at SPS,
we extrapolate down from RHIC energy by assuming that $dN_s/dy \propto
dN/dy$.

PCM results for $c$ and $b$ production in nuclear collisions \cite{rgscb}
may be overestimates, especially for lower energies, because of the
``flavor excitation'' processes that are included.  PCM results for charm
production in pp collisions match perturbative results,\cite{rmnr} if
only flavor creation is included but not flavor excitation.\cite{rgpc}
However, flavor excitation is included in PCM simulations of nuclear
collisions, so they may overestimate the number of heavy quarks that are
produced.  This overestimate is relatively small at LHC, where the two
production mechanisms have approximately equal weight, but is large below
LHC energy, where flavor excitation dominates.

Using PCM results without flavor excitation from Fig.~4 of Ref.~\cite{rgscb},
$\left. dN_c/dy\right|_{y=0} \approx 2$ for central Au+Au collisions at RHIC
and 60 at LHC.\cite{rnf7}  Simple superposition of perturbative pp collision
predictions,\cite{rmnr} assuming that $dN_c/dy$ scales as $A^{4/3}$ and that
the $c$ quarks are spread over 4 units of rapidity at RHIC and 5 units of
rapidity at LHC, gives respectively 3 and 10.  These rapidity intervals are
compatible with PCM results, but smaller than those predicted for pp
collisions.\cite{rPQCD2}

Even without flavor excitation, the PCM predicts a large increase in charm
production (compared to simple superposition of pp collisions) in Au+Au
collisions at LHC but not at RHIC.  The estimated thermal charm production
\cite{ras} is $\left. dN_c/dy\right|_{y=0} \simeq 2$ at RHIC and 10 at LHC,
which is too small to account for all of the excess charm at LHC.  We make a
conservative estimate, taking perturbative predictions \cite{rPQCD2} and
scaling $dN_c/dy$ as $A^{4/3}$, to generate the numbers in Table~2.
For $b$ quarks, we obtain all rapidity densities by taking perturbative QCD
predictions for p\=p collisions \cite{rPQCD2} and assuming that $dN_b/dy$
scales as $A^{4/3}$ (simple superposition).\cite{rscaling}

\subsection{Predictions for $Q\overline{Q}$ meson production}

Thermal recombination models have been used to analyze strange particle
production data from ultrarelativistic nuclear
collisions,\cite{rrcss,rlthsr} so we only briefly discuss mesons
containing $s$ quarks, and instead concentrate on $c$ and $b$ quark mesons.
The only significant difference between our model and others in current use
\cite{rrcss,rlthsr,rzl} is the inclusion of the pair chemical potential,
which adds an extra degree of freedom to the models.  However, we expect
that $\rho_{s\overline{s}} \simeq \rho_s \rho_{\overline{s}}$ in nearly
all collisions, so the addition of this extra degree of freedom will give
little change to the predictions of the model, while reducing its
predictive power.  Thus, while the use of $\mu_{s\overline{s}}$ may be
justified theoretically, it is probably not worthwhile including it when
analyzing collision data.  Other models also often include a baryon
chemical potential, which could be added to our model in the obvious
manner.

We show predictions for $f_Q$, the fraction of quarks of flavor $Q$ that
form $Q\overline{Q}$ mesons, in Table~III.  We include all confirmed mesons
and baryons, as well as those unconfirmed mesons that have masses less
than the highest confirmed meson mass for each flavor.\cite{rPDG}  For
simplicity, we take $\mu_s=0$.  The uncertainties in $f_c$ and $f_b$ are
very large, as they depend very strongly on $\nu$ and $T_c$.

For fixed $\nu$ and $T_c$, both $f_c$ and $f_b$ drop from S+S to Au+Au
collisions at SPS.  The magnitude of the drop in $f_b$ is largest for large
$T_c$ and large $\nu$, and could provide a constraint on the two quantities as
decrease varies from a factor of 1.2 to 4.5.  The decrease in $f_c$ could
provide a second, tighter, constraint, as $f_c$ is between 3 and 30 times
smaller in Au+Au collisions than in S+S collisions at SPS.

The energy dependences of $f_b$ and $f_c$ are also interesting.  For all
cases considered here, $f_b$ decreases monotonically with increasing energy.
The behavior of $f_c$ is non-monotonic, so changes are smaller and more
theory-dependent.  Consequently, $f_c$ will probably provide useful
constraints on $\nu$ and $T_c$ only if $\nu \simeq 10$ and $T_c \simeq 150$
MeV, in which case $f_c$ increases from SPS to LHC.

\section{Conclusions} \label{sc}

In nuclear collisions at SPS, RHIC, and LHC, $s$ and $c$ quarks are most
likely to freeze out during the period of two-phase coexistence (at the
QCD transition temperature, $T_c$) if $\nu$ is large corresponding to a
strong QCD phase transition. For moderate values of $\nu$ all heavy quark
species are most likely to freeze out from hadronic matter. The $b$ quarks
freeze out at the lowest temperatures. Since these estimates depend mainly
on the temperature dependence of the entropy density, but are independent
of the exact shape, we expect similar conclusions to hold even if the
transition is not first order as long as there is a reasonably sharp change
in the entropy density. The mean number of collisions per quark is
essentially flavor-independent, and large enough to justify the use of
quasi-equilibrium models in heavy ion collisions at and above SPS
energy.   Quasi-equilibrium models are only marginally self-consistent
for S+S collisions at SPS, and are not expected to work for pp collisions
at any attainable energy.

In the absence of exotic multi-heavy baryons, density ratios for the
various $c$ and $b$ resonances will constrain the local freezeout
densities, $\rho^{(2)}_{Q\overline{Q}}/\rho_Q$, but not the global
densities, $\rho_Q$.  Thus, determination of $\rho_c$ and $\rho_b$ may
require measurement of heavy to light resonance ratios.  The local and
global densities are predicted to be almost equal for $s$ quarks in all
collisions considered, and for $c$ quarks in collisions at LHC.  The local
$b$ quark density is much higher than the global density in all collisions
considered here.

Measurements of the fractions of $c$ and $b$ quarks forming $c\overline{c}$
and $b\overline{b}$ mesons, respectively $f_c$ and $f_b$, may help to
constrain $\nu$ and $T_c$. This appears particularly promising in light of
the strong dependence of all other quantities on the entropy density ratio
and on the critical temperature.

\acknowledgements

We thank P. L{\'{e}}vai, L. McLerran, and R. Vogt for stimulating
discussions. This work was supported in part by the U.S. Department of Energy
under Grant No.\ DOE/DE-FG02-86ER-40251.

\begin{table}[tbh]

\caption{Freezeout conditions for $s$, $c$, and $b$ quarks.  A ``c'' after
$T_f$ indicates that freezeout was due to loss of thermal contact, while
``w'' denotes freezeout due to emission.}

\vspace{12pt}
\begin{center}
\begin{tabular}{lrr cdddd cdddd}
 & & && \multicolumn{4}{c}{$T_f$ (MeV)} &&
  \multicolumn{4}{c}{$\tau_f$ (fm/$c$)} \\
 \multicolumn{3}{l}{$T_c$ (MeV):} && \multicolumn{2}{c}{150} &
  \multicolumn{2}{c}{200} && \multicolumn{2}{c}{150} &
  \multicolumn{2}{c}{200} \\
 $Q$ & $A$ & $dN/dy$ && $\nu$:~5 & 10 & 5 & 10 && 5 & 10 & 5 & 10 \\ \hline
 $s$ &  32 &   85 && 150c & 150c & 198c & 200c && 2.6 & 2.1 & 2.0 & 1.3 \\
     & 197 & 1000 && 145w & 150w & 162c & 200c &&  18 &  16 &  13 &  12 \\
     & 197 & 2000 && 150w & 150w & 176w & 200w &&  23 &  22 &  21 &  18 \\
     & 197 & 3500 && 150w & 150w & 197w & 200w &&  28 &  27 &  25 &  23 \\
 \hline
 $c$ &  32 &   85 && 140c & 150c & 176c & 200c && 5.8 & 3.8 & 2.9 & 2.0 \\
     & 197 & 1000 && 116w & 150w & 138c & 182c &&  36 &  30 &  21 &  18 \\
     & 197 & 2000 && 134w & 150w & 149w & 198w &&  46 &  39 &  33 &  29 \\
     & 197 & 3500 && 150w & 150w & 168w & 200w &&  55 &  49 &  41 &  35 \\
 \hline
 $b$ &  32 &   85 && 110c & 145c & 138c & 182c &&  12 &  10 & 6.0 & 5.2 \\
     & 197 & 1000 &&  91w & 120w & 108c & 143c &&  75 &  65 &  44 &  38 \\
     & 197 & 2000 && 104w & 138w & 117w & 154w &&  98 &  84 &  70 &  60 \\
     & 197 & 3500 && 117w & 150w & 131w & 173w && 121 & 103 &  87 &  75 \\
 \hline
 & & && \multicolumn{4}{c}{$r_Q$ (fm)} && \multicolumn{4}{c}{$n_f$} \\ \hline
 $s$ &  32 &   85 && 2.9 & 2.5 & 2.6 & 1.9 &&  1 &  1 &  1 &  1 \\
     & 197 & 1000 && 9.3 & 9.3 & 8.2 & 8.5 &&  5 &  4 &  4 &  3 \\
     & 197 & 2000 && 9.3 & 9.3 & 9.3 & 9.3 &&  8 &  7 &  6 &  5 \\
     & 197 & 3500 && 9.3 & 9.3 & 9.3 & 9.3 && 11 & 11 &  9 &  8 \\ \hline
 $c$ &  32 &   85 && 3.2 & 2.5 & 2.4 & 2.0 &&  1 &  1 &  1 &  1 \\
     & 197 & 1000 && 9.3 & 9.3 & 8.0 & 8.1 &&  6 &  5 &  4 &  3 \\
     & 197 & 2000 && 9.3 & 9.3 & 9.3 & 9.3 &&  9 &  8 &  7 &  6 \\
     & 197 & 3500 && 9.3 & 9.3 & 9.3 & 9.3 && 13 & 12 & 10 &  8 \\ \hline
 $b$ &  32 &   85 && 3.1 & 3.2 & 2.3 & 2.3 &&  2 &  1 &  1 &  1 \\
     & 197 & 1000 && 9.3 & 9.3 & 7.9 & 7.9 &&  7 &  4 &  4 &  4 \\
     & 197 & 2000 && 9.3 & 9.3 & 9.3 & 9.3 && 13 & 11 &  8 &  7 \\
     & 197 & 3500 && 9.3 & 9.3 & 9.3 & 9.3 && 19 & 16 & 13 & 11
 \end{tabular}
 \end{center}
\end{table}

\begin{table}[tbh]

\caption{Freezeout densities.}

\vspace{12pt}
\begin{center}
\begin{tabular}{lrrdcddddcdddd}
 & & & && \multicolumn{4}{c}{$\rho_Q$ (10$^{-5}$ fm$^{-3}$)} &&
  \multicolumn{4}{c}{$\rho^{(2)}_{Q\overline{Q}}/\rho_Q\rho_{\overline{Q}}$} \\
 \multicolumn{4}{l}{$T_c$ (MeV):} && \multicolumn{2}{c}{150} &
  \multicolumn{2}{c}{200} && \multicolumn{2}{c}{150} &
  \multicolumn{2}{c}{200} \\
 $Q$ & $A$ & $dN/dy$ & $dN_Q/dy$ && $\nu$:~5 & 10 & 5 & 10 &&
  5 & 10 & 5 & 10 \\ \hline
 $s$ &  32 &   85 & 5      && 4200 & 5300 & 5400 & 8600 &&
                               1.1 &  1.1 &  1.1 &  1.2 \\
     & 197 & 1000 & 25     &&  890 & 1000 & 1200 & 1300 &&
                               1.0 &  1.0 &  1.0 &  1.0 \\
     & 197 & 2000 & 50     && 1400 & 1500 & 1600 & 1800 &&
                               1.0 &  1.0 &  1.0 &  1.0 \\
     & 197 & 3500 & 250    && 5800 & 6000 & 6400 & 7200 &&
                               1.0 &  1.0 &  1.0 &  1.0 \\ \hline
 $c$ &  32 &   85 & 0.03   &&   11 &   18 &   22 &   33 &&
                                30 &   41 &   37 &   45 \\
     & 197 & 1000 & 0.3    &&  5.5 &  6.4 &  9.2 &   11 &&
                               3.5 &  3.1 &  3.4 &  3.0 \\
     & 197 & 2000 & 1      &&   14 &   17 &   20 &   23 &&
                               2.0 &  1.8 &  1.7 &  1.6 \\
     & 197 & 3500 & 5      &&   59 &   66 &   79 &   93 &&
                               1.2 &  1.2 &  1.2 &  1.1 \\ \hline
 $b$ &  32 &   85 & 0.0003 && 0.06 & 0.06 & 0.11 & 0.13 &&
                              6700 & 5600 & 8100 & 6800 \\
     & 197 & 1000 & 0.003  && 0.03 & 0.03 & 0.05 & 0.05 &&
                               530 &  450 &  500 &  430 \\
     & 197 & 2000 & 0.02   && 0.13 & 0.16 & 0.19 & 0.22 &&
                               100 &   89 &   75 &   65 \\
     & 197 & 3500 & 0.4    &&  2.2 &  2.5 &  3.0 &  3.5 &&
                               7.4 &  6.4 &  5.6 &  5.0
 \end{tabular}
 \end{center}
\end{table}

\begin{table}[tbh]

\caption{Predicted $c\overline{c}$ and $b\overline{b}$ meson fractions.}

\vspace{12pt}
\begin{center}
\begin{tabular}{rrcddddcdddd}
 & && \multicolumn{4}{c}{$f_c$} && \multicolumn{4}{c}{$f_b$} \\
 \multicolumn{2}{l}{$T_c$ (MeV):} &&
 \multicolumn{2}{c}{150} & \multicolumn{2}{c}{200} &&
 \multicolumn{2}{c}{150} & \multicolumn{2}{c}{200} \\
 $A$ & $dN/dy$ && $\nu$:~5 & 10 & 5 & 10 && 5 & 10 & 5 & 10 \\ \hline
  32 &      85 && 0.066 & 0.075  & 0.029 & 0.021  &&
                  0.62  & 0.16   & 0.36  & 0.072  \\
 197 &    1000 && 0.021 & 0.002  & 0.008 & 0.001  &&
                  0.50  & 0.053  & 0.20  & 0.016  \\
 197 &    2000 && 0.009 & 0.004  & 0.004 & 0.001  &&
                  0.20  & 0.013  & 0.070 & 0.005  \\
 197 &    3500 && 0.009 & 0.010  & 0.005 & 0.002  &&
                  0.075 & 0.008  & 0.027 & 0.002
 \end{tabular}
 \end{center}
\end{table}

\vfill \eject


\begin{references}
\bibitem{rnf7} We do not use the information contained in Fig.~7 of
Ref.~\cite{rgscb}, as this figure is incorrect.\cite{rgpc}
\bibitem{rscaling} This scaling is different from PCM predictions.
The PCM scaling exponents for $dN_Q/dy$ in p$A$ collisions, $\beta_Q$, are
approximately equal to the observed scaling exponents for $d\sigma_Q/dy$,
$\alpha_Q$.  However, $dN_Q/dy = (1/\sigma_{inelastic})d\sigma_Q/dy$, and
the inelastic cross section $\sigma_{inelastic} \propto A^{2/3}$, so
because $\beta_Q \gg \alpha_Q-2/3$ the PCM prediction is not supported by
data.
\end{references}
\end{document}